\documentclass[conference]{IEEEtran}
\IEEEoverridecommandlockouts
\usepackage{amsmath,amssymb,amsfonts}
\usepackage{algorithmic}
\usepackage{graphicx}
\usepackage{textcomp}
\usepackage{xcolor}
\usepackage{biblatex} 
\usepackage{booktabs}
\usepackage{bbm}
\usepackage{multirow}
\usepackage{subfiles}
\usepackage{bm}
\usepackage[normalem]{ulem}

\def\BibTeX{{\rm B\kern-.05em{\sc i\kern-.025em b}\kern-.08em
    T\kern-.1667em\lower.7ex\hbox{E}\kern-.125emX}}

\newcommand{\ok}{\color{black}}

\begin{document}

\title{Bilateral {\ok Unsymmetrical} Graph Contrastive Learning for Recommendation\\
{\footnotesize }
\thanks{*Corresponding author}
}

\author{\IEEEauthorblockN{Anonymous Authors}}

\author{\IEEEauthorblockN{1\textsuperscript{st} Jiaheng Yu}
\IEEEauthorblockA{
\textit{School of Computer Science} \\
\textit{Wuhan University}\\
Wuhan, China \\
yujiaheng@whu.edu.cn}

\\

\IEEEauthorblockN{4\textsuperscript{th} Kai Zhu}
\IEEEauthorblockA{
\textit{School of Computer Science} \\
\textit{Wuhan University}\\
Wuhan, China \\
zhukai-cs@whu.edu.cn}

\\
\and

\IEEEauthorblockN{2\textsuperscript{nd} Jing Li*}
\IEEEauthorblockA{
\textit{School of Computer Science} \\
\textit{Wuhan University}\\
Wuhan, China \\
leejingcn@whu.edu.cn}

\\

\IEEEauthorblockN{5\textsuperscript{th} Shuyi Zhang}
\IEEEauthorblockA{
\textit{School of Computer Science} \\
\textit{Wuhan University}\\
Wuhan, China \\
2020300004062@whu.edu.cn}

\\
\and

\IEEEauthorblockN{3\textsuperscript{rd} Yue He}
\IEEEauthorblockA{
\textit{School of Computer Science} \\
\textit{Wuhan University}\\
Wuhan, China \\
yuehe.cs@whu.edu.cn}

\\

\IEEEauthorblockN{6\textsuperscript{th} Wen Hu}
\IEEEauthorblockA{
\textit{School of Artificial Intelligence} \\
\textit{Wuchang University of Technology}\\
Wuhan, China \\
huwen@wut.edu.cn}
}

\maketitle

\begin{abstract}

Recent methods utilize graph contrastive Learning within graph-structured user-item interaction data for collaborative filtering and have demonstrated their efficacy in recommendation tasks. However, they ignore that the difference relation density of nodes between the user- and item-side causes the adaptability of graphs on bilateral nodes to be different after multi-hop graph interaction calculation, which limits existing models to achieve ideal results. To solve this issue, we propose a novel framework for recommendation tasks called \uline{B}ilateral \uline{U}n\uline{s}ymmetrical \uline{G}raph \uline{C}ontrastive \uline{L}earning (BusGCL) that consider the bilateral unsymmetry on user-item node relation density for sliced user and item graph reasoning better with bilateral slicing contrastive training. 
Especially, taking into account the aggregation ability of hypergraph-based graph convolutional network (GCN) in digging implicit similarities is more suitable for user nodes, embeddings generated from three different modules: hypergraph-based GCN, GCN and perturbed GCN, are sliced into two subviews by the user- and item-side respectively, and selectively combined into subview pairs bilaterally based on the characteristics of inter-node relation structure. Furthermore, to align the distribution of user and item embeddings after aggregation, a dispersing loss is leveraged to adjust the mutual distance between all embeddings for maintaining learning ability. Comprehensive experiments on two public datasets have proved the superiority of BusGCL in comparison to various recommendation methods. Other models can simply utilize our bilateral slicing contrastive learning to enhance recommending performance without incurring extra expenses.

\end{abstract}


\section{Introduction} 
\label{sec:int}
Recommendation systems have found widespread application in diverse domains, including online retail platforms \cite{Huang2019OnlinePP}, social networking applications, and online multimedia websites \cite{9216015}, to aid users in navigating through the overwhelming amounts of information on the internet and discover items that align with their preferences. However, recommending tasks remains difficult because of the distinct structure of the data and the extremely sparse density of the user-item dataset.


\begin{figure}[t]
\centerline{\includegraphics[scale=0.5]{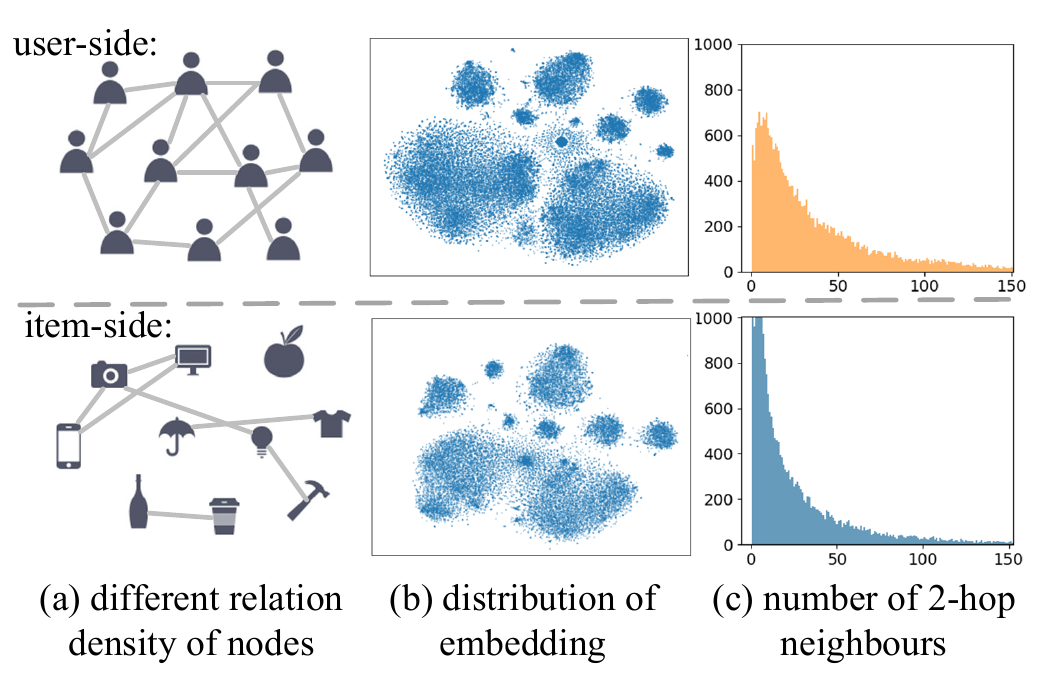}}
\caption{
 Illustration of difference in the relation density between bilateral nodes. (a) means that user nodes often have denser inter-node relationships than item nodes. (b) visualizes the different distribution of embeddings generated from a 1-layer LightGCN \cite{LightGCN-He2020LightGCNSA} on Yelp dataset by t-SNE.  (c) counts the number of 2-hop neighbours, by normalizing them into a same total number. 
 }



\label{fig:tsne}
\end{figure}

Based graph convolutional network (GCN) methods for recommendation \cite{Salakhutdinov2007ProbabilisticMF,He2016FastMF} consider collaborative filtering as the fundamental architecture, that reduces the dimensionality of users and items by projecting observed interactions onto a low dimensional space for representation \cite{Dong2017AHC}.
Further exploiting the hypergraph structures, hypergraphs show their potential in accurately representing more implicit high-order relationships within graph data \cite{HCCF-Xia2022HypergraphCC}. Hypergraph-based recommenders \cite{HCCF-Xia2022HypergraphCC} and \cite{SHT-Xia2022SelfSupervisedHT} utilize hyperedges to encapsulate implicit high-order collaborative effects within user-item graphs. Recently, combining the self-supervised Graph Contrastive Learning (GCL) paradigm with strong effectiveness in resisting data sparsity. Subsequently, several recommendation models based on hypergraph structures fusing with GCL \cite{HCCF-Xia2022HypergraphCC,SHT-Xia2022SelfSupervisedHT,Yu2021SelfSupervisedMH} have been proposed and led a promising development of recommendation. However, these symmetrical GCL-based models generally overlooked the inter-node pattern differences upon the interaction information from the perspectives of users and items, which could lead to the differences in probability distribution of the learned embeddings. 

The difference in relation density of nodes between the user- and item-side is shown in Fig. \ref{fig:tsne} (a), which represents that user nodes often have denser inter-node relationships than item ones and are more inclined to organize groups of similarity. For example, after GCN reasoning, the different distributions of embeddings generated from a 1-layer LightGCN \cite{LightGCN-He2020LightGCNSA} on Yelp \cite{Yelp-Yin2019SocialIG} by t-SNE(b) are visualized in Figure \ref{fig:tsne}(b). Comparing the two figures, it can be seen that embeddings of users are more cohesive with more clear boundaries of groups. That means similar users generally have closer relationships. Then, different relational density brings different degrees of aggregation in graph structures after multi-hop interaction calculation. Before the 2-hop calculation, Fig. \ref{fig:tsne}(c) shows counts of the number of 2-hop neighbors of the user node and the item node, that statistic is normalizing into the same total number. It shows that item nodes are concentrated in the left end representing fewer complex relationships, and user nodes are relatively balanced. Different relational density brings different degrees of aggregation in graph structures after multi-hop interaction calculation. In view of this, incorporating identical or highly similar graph structures without differentiated methods on both user-side and item-side takes no effective measure to address this situation, limiting improvement in recommending performance.

To tackle this limitation, we propose a novel framework for recommendation system, namely \uline{B}ilateral \uline{U}n\uline{s}ymmetrical \uline{G}raph \uline{C}ontrastive \uline{L}earning (BusGCL), which consider the bilateral unsymmetry on user-item node relation density for sliced user and item graph reasoning better with bilateral slicing contrastive learning. There is a multi-structure graph model to extract the sliced view of users using hypergraph-approach and of items using the perturbing-based GCN method, which allows for the construction of more effective and expressive contrastive views. In theoretical analysis, the hyperedges of hypergraphs tend to aggregate nodes with similar relation patterns, which is more suitable for user-side nodes with widespread inter-node similarity, while the features of item-side nodes are generally scattered. Thus, we adapt GCN with random noise perturbing to capture collaborative information on the item-side to generate contrastive views. Furthermore, in order to mitigate the over-smooth issue induced by the introduction of noise perturbing, we designed dispersing loss to balance it, thereby maintaining the learning ability of nodes to refine the collaborative relationship information during training. In summary, the contribution of this work is threefold:

\begin{itemize}
\item {\ok   We enhance the recommendation system by utilizing the bilateral unsymmetry of node density on the user- and item-side,  and propose bilateral slicing contrastive learning which generates user and item subviews through different GCNs to reason better results.  } 

\item {\ok  We propose a multi-struct graph framework BusGCL considering the characteristics of different gcns. BusGCL provides guidance for other recommendation methods to utilize hypergraphs in user-side aggregation. } 


\item {\ok A dispersing loss is designed to alleviate the over-smoothing issue deteriorated by GCN, and it refines bilateral slicing contrastive training. The outperforming results on the experiments of different datasets illustrate the efficiency of our model. }


\end{itemize}


\section{Related Work}
\label{sec:related}

\subsection {Recommendation methods}


The fundamental premise underpinning numerous collaborative filtering models \cite{Konstan1997GroupLensAC,Resnick1994GroupLensAO,Sarwar2001ItembasedCF}. making recommendation for the target user by finding other users who are similar to the target user or other items that are similar to the target item. However, recent recommenders based on collaborative filtering extend to three different types: \textbf{Graph Convolution Networks based Recommenders.}  NGCF \cite{NGCF-10.1145/3331184.3331267} is a graph-based collaborative filtering method that integrates features of second-order interactions into the messages during the message-passing process. LightGCN \cite{LightGCN-He2020LightGCNSA} designs a lightweight graph convolution for training efficiency and generation ability with only adding neighborhood aggregation as a component. \textbf{Hypergraph-based recommenders}.HCCF \cite{HCCF-Xia2022HypergraphCC} enhances GNN with hypergraph learning global dependency, and employs cross-view contrastive learning to capture both local and global collaborative relationships simultaneously. SHT \cite{SHT-Xia2022SelfSupervisedHT} introduces transformer architecture into the hypergraph recommendation to improve recommendation performance. \textbf{Self-Supervised Learning enhanced recommenders}. SLRec \cite{Yao2020SelfsupervisedLF} incorporates contrastive learning between features to regularize two augmented embeddings, in order to enhance the effectiveness of data augmentation based recommendation. And SGL \cite{SGL-Wu2020SelfsupervisedGL} generates contrastive views through three different ways as node dropout, edge dropout and random walk to enhance recommendation performance through contrastive learning. Recently, SimGCL \cite{SimGCL-Yu2021AreGA} refines the graph augmentation procedure within contrastive learning by directly incorporating noises taking values in hypersphere space randomly.

\subsection{ Graph Contrastive Learning in Recommendation}

Contrastive learning has attracted widespread attention in computer vision \cite{10144391}, which constructs positive and negative sample pairs through differences between views to provide a self-supervised solution to data sparsity problem \cite{Yu2018AdaptiveIF,Yu2019GeneratingRF}. Inspired by contrastive learning, S3-Rec \cite{Zhou2020S3RecSL} firstly employs random mask on attributes, sequences and items, thereby generating sequence augmentations for the pre-training of sequential models through the maximization of mutual information with contrastive learning. Beyond the application of the dropout, CL4Rec \cite{Xie2022ContrastiveLF} suggests the reordering and cropping of item segments for sequential data augmentation.

In addition to addressing the data sparsity problem, CLRec \cite{Zhou2020ContrastiveLF} has theoretically demonstrated that contrastive learning can also alleviate the exposure bias present in recommendations, and improve the depth of matching with respect to fairness and efficiency. SGL \cite{SGL-Wu2020SelfsupervisedGL} employs node/edge drop techniques, coupled with random walk methods to generate positive instances. HCCF \cite {HCCF-Xia2022HypergraphCC} uses hypergraph to generate contrastive view for high-order collaborative signals learning in intereaction graph and achieve notable success. 
{\ok However, these methods seldom consider the difference relation density of nodes between the user-side and item-side. It is still a problem that they deal with the whole embedding which contains different distributed user- and item embedding to get loss after the GCN aggregation, which limits the model not getting a great result. }


\section{Methodology}
\label{sec:method}

\begin{figure*}[t]
\centerline{\includegraphics[scale=0.60]{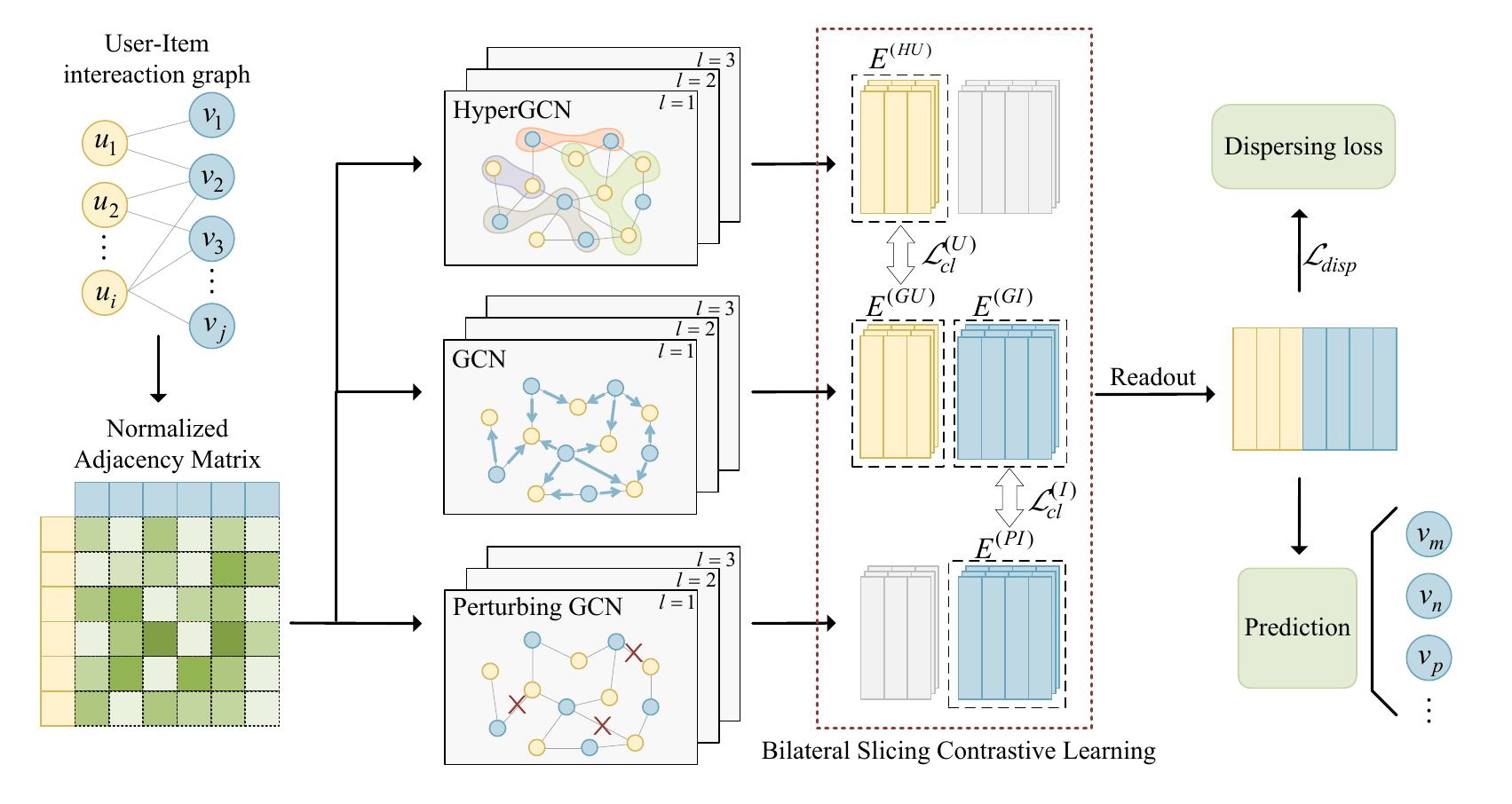}}
\caption{{\ok Illustration of BusGCL framework.  An adjacency matrix which represents user-item interaction graph, passes through a Multi-structurally Graph Model that contains three variants of GCNs to form three embedding matrices. And three embedding matrices are sliced by both user-side and item-side for bilateral slicing contrastive learning before recommendation predictions.}}

\label{fig:struct}
\end{figure*}

\subsection{Overview}
\label{sec:pre}
Perform mathematical expression in preliminaries, we represent the sets of users and items as \(\mathcal{U}={{u_1,u_2,\cdots,u_i}},(|\mathcal{U}|=I)\) and \(\mathcal{V}={{v_1,v_2,\cdots,v_j}},(|\mathcal{V}|=J)\), respectively. The interaction adjacency matrix \(\mathcal{A}\in\bm{R}^{I\times{J}}\) stores interaction history information between users \(\mathcal{U}\) and corresponding items \(\mathcal{V}\). The value of each entry \(\mathcal{A}_{i,j}\) in \(\mathcal{A}\) is designated as \(1\) when there exists an interaction between the user \(u_i\) and the item \(v_j\), and \(\mathcal{A}_{i,j}=0\) otherwise.

The comprehensive architecture of BusGCL is depicted in Figure \ref{fig:struct}. First, we get better user-item interaction representations from our multi-structurally graph Model. When inputting user-item interaction graph to the model, an adjacency matrix \(\mathcal{A}\) is constructed with aggregation, followed by the extraction of representations via three variants of GCNs to form three embedding matrices  $\bm{E}^{(G)},\bm{E}^{(P)},\bm{E}^{(H)}$. Then, we use bilateral slicing contrastive learning to realize great recommendation. There embedding matrices are sliced by both user-side and item-side, and are recombined into two unsymmetrical subview pairs for contrastive learning. The structure of embeddings is then constrained by dispersing loss, culminating in the generation of recommendation predictions. 


\subsection{A Multi-structurally Graph Model for Recommendation}
The Multi-structurally Graph Model is structurally consist of two steps: Obtaining based Adjacency Matrix and multi-structural graph reasoning.

\textbf{Obtaining based Adjacency Matrix.} Firstly, we normalize the adjacency matrix \(\mathcal{A}\) mentioned above which encapsulates the interaction relation between users and items, represented by the subsequent formula:

\begin{equation}
\label{eq:adj}
\Bar{\mathcal{A}}=\bm{D}_{(u)}^{-1/2}\cdot\mathcal{A}\cdot\bm{D}_{(v)}^{-1/2},
\Bar{\mathcal{A}}_{i,j}=\cfrac{\mathcal{A}_{i,j}}{\sqrt{|\mathcal{N}_i|\cdot|\mathcal{N}_i|}} ,
\end{equation}
here, \(\bm{D}_{(u)}\in\mathbbm{R}^{I{\times}I}\),\(\bm{D}_{(v))}\in\mathbbm{R}^{J{\times}J}\) represents degree matrices of users and items, respectively. \(\mathcal{N}_i\) represents neighbouring item nodes of user \(u_i\), \(\mathcal{N}_j\) is similar. Then, to encode the pattern information of user-item interactions, we follow the classic conventional collaborative filtering paradigm, projecting the graph structure into a \(d\)-dimensional latent space. For user \(u_i\) and item \(v_j\), we establish vectors \(\bm{e}_i\) and \(\bm{e}_j\) of size \(\mathbbm{R}^d\) as embeddings, and define matrices combined by embeddings as \(\bm{E}^{(u)}\in\mathbbm{R}^{I{\times}d}\) and  \(\bm{E}^{(v)}\in\mathbbm{R}^{J{\times}d}\), respectively. 

\textbf {Multi-structural graph reasoning.} 
We use a three-branch graph model to reasoning with inputted embedding \(\bm{E}^{(u)}\) and \(\bm{E}^{(v)}\).
The normal branch is the middle branch shown in Figure \ref{fig:struct}. To aggregate the local collaborative signals for each node from their neighbours, Simplified by LightGCN \cite{LightGCN-He2020LightGCNSA}, we design an embedding propagation layer leveraging a lightweight graph convolutional network without non-linear activation functions. And the output embedding $\bm{E}_l^{(G)}=\{\bm{E}_l^{(GU)} ; \bm{E}_l^{(GI)} \}$ from $l$-th layer network contains user part $\bm{E}_l^{(GU)} = \{ \bm{\alpha}_{1,l}^{(u)}, \bm{\alpha}_{2,l}^{(u)},... \bm{\alpha}_{i,l}^{(u)} |i \in I\} $ and item part $\bm{E}_l^{(GI)} = \{ \bm{\alpha}_{1,l}^{(v)}, \bm{\alpha}_{2,l}^{(v)},... \bm{\alpha}_{j,l}^{(v)} | j \in J \} $, which $I$ and $J$ denote the number of user and item respectively. This process can be described as follows:

\begin{equation}
    \label{eq:gcnlayer}
    \bm{\alpha}_{i,l}^{(u)}=\Bar{\mathcal{A}}_{i,*}\cdot\bm{E}_{l-1}^{(G)},
    \bm{\alpha}_{j,l}^{(v)}=\Bar{\mathcal{A}}_{*,j}\cdot\bm{E}_{l-1}^{(G)},
\end{equation}
where \(\bm{\alpha}_i^{(u)},\bm{\alpha}_j^{(v)}\in\mathbbm{R}^d\) presents the aggregated collaborative information of centric nodes. 


In order to refine multi-hop neighbours' relation, we integrate multiple embedding propagation layers as a graph neural network. Combining residual connection to avoid gradient vanishing \cite{He2015DeepRL}, we operate the Readout on different layers to get embeddings $\bm{e}_{i,l}^{(u)} \in  \bm{\Bar{E}}_l^{(GU)} $ and $ \bm{e}_{j,l}^{(v)} \in \bm{\Bar{E}}_l^{(GI)} $, in which embedding of $l$-th layer is used to predict next user-item relation. Neighbour information transmission is towards the following formula:
\begin{equation}
    \label{eq:gcnemb}
    \bm{e}_{i,l}^{(u)}=\bm{e}_{i,l-1}^{(u)}+\bm{\alpha}_{i,l}^{(u)},
    \bm{e}_{j,l}^{(v)}=\bm{e}_{j,l-1}^{(v)}+\bm{\alpha}_{j,l}^{(v)}.
\end{equation}


The hyperedges of hypergraph \cite{HCCF-Xia2022HypergraphCC} can connect any number of vertices, forming a similar effect of complete subgraphs with weighted attributes on the interaction graph, which is beneficial for aggregating non-adjacent but potentially similar nodes by leveraging hyperedges as intermediate hubs. In the top branch in Figure \ref{fig:struct}, based on the embedding result $ \bm{E}_l^{(G)} $ from GCN reasoning in middle branch, the output embedding $\bm{E}_l^{(H)}$ which contains two hypergraphs is defined with $H$ hyperedges to represent users and items as $\mathcal{H}^{(u)}=\{ \gamma_1^{(u)}, \gamma_2^{(u)},...\gamma_l^{(u)} \} \in \mathbbm{R}^{I \times H} $ and $\mathcal{H}^{(v)}= \{ \gamma_1^{(v)}, \gamma_2^{(v)},...\gamma_l^{(v)} \}\in\mathbbm{R}^{J \times H} $. The progress is as follow:
\begin{equation}
    \label{eq:hyperlayer}  \bm{\gamma}_l^{(u)}=LeakyReLU(\mathcal{H}^{(u)}\cdot\mathcal{H}^{(u)\top}\cdot\bm{E}_{l-1}^{(G)}).
\end{equation}
Hyper embeddings of items $\gamma_l^{(v)}$ can be derived following a similar way, where $\mathcal{H}^{(u)} $ and $\mathcal{H}^{(v)} $ are substituted by a low-rank approximation to reduce computational cost, which is computed with $l$-th readout embedding follows:
\begin{equation}
\label{eq:hyperadj}
{\mathcal{\hat{H}}_l^{(u)}}=\bm{\Bar{E}}_l^{(GU)}\cdot\bm{W}^{(u)},
\mathcal{\hat{H}}_l^{(v)}=\bm{\Bar{E}}_l^{(GI)}\cdot\bm{W}^{(v)},
\end{equation}
where \(\bm{W}^{(u)},\bm{W}^{(v)} \in \mathbbm{R}^{d{\times}H}\) are the parameter-learnable matrices representing the hyperedges for users and items. 


In the bottom branch in Figure \ref{fig:struct}, following SimGCL \cite{SimGCL-Yu2021AreGA}, we adapt an another GCN which adds imperceptibly small perturbation. At each layer, the current embeddings are perturbed by a stochastic noise \(\Delta\), this corresponds to a numerical equivalence with points located on a hypersphere of a given radius \(r\) . ${||\Delta||}_2=r $, and $\Delta = \Bar{\Delta} {\odot} sign{({\bm{e})}}, \Bar{\Delta}\in\mathbbm{R}^d\sim{U(0,1)} $. Similar to the form of the above equation \ref{eq:gcnlayer}, the embeddings $ {E}_l^{(P)}$ which is consist of ${E}_l^{(PU)}= \{ \beta_{1,l}^{(u)}, \beta_{2,l}^{(u)}, ...\beta_{i,l}^{(u)} |i \in I\}$ and $\bm{E}_l^{(PI)} = \{{\beta}_{1,l}^{(u)}, {\beta}_{2,l}^{(u)},...{\beta}_{j,l}^{(u)}|j \in J\}$ obtained from $l$-th layer in the perturbing-GCN follows:
\begin{equation}
    \label{eq:perturblayer}
    \bm{\beta}_{i,l}^{(u)}=\Bar{\mathcal{A}}_{i,*}\cdot\bm{E}_{l-1}^{(P)}+\Delta',
    \bm{\beta}_{j,l}
    ^{(v)}=\Bar{\mathcal{A}}_{*,j}\cdot\bm{E}_{l-1}^{(P)}+\Delta'' .
\end{equation}


\subsection{Bilateral Slicing Contrastive Learning}   
After a \(L\)-layer propagating progression of three kinds of GCN introduced above (GCN, GCN with perturbing and HyperGCN), we stack the outputs of each layer and obtain three matrices of embeddings with isomorphic structures and complementary semantics denote as \(\bm{E}^{(G)},\bm{E}^{(P)},\bm{E}^{(H)} \in \mathbbm{R}^{(I+J){\times}d{\times}L}\), respectively. These matrices can be recognized as views for contrastive learning because of the implicit supervising signals derived from subtle differences in latent space. Considering the inter-node distribution difference between bilateral nodes discussed in Section \ref{sec:int}, we slice each view into two subviews by side as illustrated in Figure \ref{fig:struct}.

\begin{table}[]
\centering
\caption{Statistics of the Experimental Datasets}
\label{tab:dataset}
\begin{tabular}{ccccc}
\toprule
Datasets    & \# Users & \# Items & \# Interactions & \ Density \\ 
\midrule
Yelp        & 42712    & 26822    & 182357          & 1.6\(e^{-4}\)    \\
Last.FM     & 1892     & 17632    & 92834           & 2.8\(e^{-3}\)    \\
\bottomrule
\end{tabular}
\end{table}

Considering that nodes on the user-side often have more relationship density, which means that user nodes have more similarity, we adopt a hypergraph structure equipped with hyperedges with node aggregation ability to model relation on this side. On the contrary, the number of neighbors of the item-side nodes is generally smaller, indicating that the features between items are relatively more scattered. On this side, we choose GCN with perturbing that combines noise disturbance and is better at learning differences between nodes. In general, we select the user-side subview from the Hypergraph-GCN \(\bm{E}^{(HU)}\) and the item-side sub-view from GCN with perturbing \(\bm{E}^{(PI)}\) to compare with the two sub-views from GCN \(\bm{E}^{(GU)}, \bm{E}^{(GI)}\), and use the InfoNCE \cite{infonce-oord2019representation} function to calculate the contrastive loss by layer as:
\begin{equation}
    \label{eq:infonce_U}
    \mathcal{L}_{cl}^{(U)}=\sum_{i=0}^I{\sum_{l=0}^L{-log\frac{exp(sim(\bm{\alpha}_{i,l}^{(u)},\bm{\gamma}_{i,l}^{(u)})/\tau_c)}{\sum_{i'=0}^{I}{exp(sim(\bm{\alpha}_{i,l}^{(u)},\bm{\gamma}_{i',l}^{(u)})/\tau_c)}}}},
\end{equation}
\begin{equation}
    \label{eq:infonce_I}
    \mathcal{L}_{cl}^{(I)}=\sum_{j=0}^J{\sum_{l=0}^L{-log\frac{exp(sim(\bm{\alpha}_{j,l}^{(v)},\bm{\beta}_{j,l}^{(v)})/\tau_c)}{\sum_{j'=0}^{J}{exp(sim(\bm{\alpha}_{j,l}^{(v)},\bm{\beta}_{j',l}^{(v)})/\tau_c)}}}},
\end{equation}
where \(sim(\cdot)\) means the cosine similarity function, while utilizing a temperature coefficient $\tau_c$ for sensitivity in calculating contrastive learning loss, and \(L\) denotes the max number of convolutional layers.

\subsection{Dispersing Loss}
Introducing noise in GCN with perturbing without limitations will cause the distribution of embedded features to tend towards over equilibrium, resulting in having relatively small distances in the latent space gradually, further blurring already subtle differences between nodes and exacerbating the phenomenon of over-smoothing.
To this end and inspired by InfoNCE which has the ability to push away negative samples in vector space, we introduce a variation of infoNCE as a metric loss function to constrain the distance of embeddings. Treating all the other vectors of a single embedding matrix as negative samples, making contrastive learning with its own view, and achieving the dispersion effect that all embeddings are gradually dispersed to maintain sufficient distance for learning knowledge. We apply this constraint on the readout of the matrices obtained by GCN:
\begin{equation}
    \label{eq:lossDisp}
\mathcal{L}_{disp}=\sum_{k=0}^{I+J}{-log\frac{exp(sim(\bm{R}_{k},\bm{R}_{k})/\tau_d)}{\sum_{k'=0}^{I+J}{exp(sim(\bm{R}_{k},\bm{R}_{k'})/\tau_d)}}},
\end{equation}
where $\bm{R}$ represents the result ${\Bar{E}}_l^{(G)}$ from the readout layer. 

\begin{table*}[]
\centering
\caption{Overall performance comparison in terms of Recall and NDCG on three datasets, where the best-performing results under each metric are shown in bold, while the second best results are highlighted with underlines.}
\label{tab:overall}
\begin{tabular}{ccccccccc}
\toprule
\multirow{3}{*}{Methods} & \multicolumn{4}{c}{Yelp} & \multicolumn{4}{c}{Last.FM} \\
\cmidrule(r){2-5}
\cmidrule(r){6-9}

         & Recall@20 $\uparrow$ & Recall@40 $\uparrow$ & NDCG@20 $\uparrow$ & NDCG@40 $\uparrow$  & Recall@20 $\uparrow$ &Recall@40 $\uparrow$ & NDCG@20 $\uparrow$ & NDCG@40 $\uparrow$ \\
\midrule
NGCF \cite{NGCF-10.1145/3331184.3331267}    & 0.0681 & 0.1019 & 0.0336 & 0.0419  & 0.2081 & 0.2944 & 0.1474 & 0.1829 \\
LightGCN  \cite{LightGCN-He2020LightGCNSA} & 0.0761 & 0.1175 & 0.0373 & 0.0474  & 0.2349 & 0.3220 & 0.1704 & 0.2022 \\
HCCF   \cite{HCCF-Xia2022HypergraphCC}  & 0.0789 & 0.1185 & 0.0399 & 0.0496  & 0.2410 & 0.3232 & 0.1773 & 0.2051 \\
SHT    \cite{SHT-Xia2022SelfSupervisedHT}  & 0.0794 & 0.1217 & 0.0395 & 0.0497  & 0.2420 & 0.3235 & 0.1770 & 0.2055 \\
SLRec  \cite{SLRec10.1145/3459637.3481952}  & 0.0665 & 0.1032 & 0.0327 & 0.0418  & 0.1957 & 0.2792 & 0.1442 & 0.1737 \\ 
SGL   \cite{SGL-Wu2020SelfsupervisedGL}   & 0.0803 & 0.1226 & 0.0398 & 0.0502  & \uline{0.2427} & \textbf{0.3405} & 0.1761 & \textbf{0.2104} \\
SimGCL \cite{SimGCL-Yu2021AreGA}  & \uline{0.0813} & \uline{0.1230} & \uline{0.0408} & \uline{0.0510}  & 0.2398 & \uline{0.3337} & \uline{0.1780} & \uline{0.2099} \\
\midrule
\textbf{BusGCL}    & \textbf{0.0840} & \textbf{0.1263} & \textbf{0.0424} & \textbf{0.0528}    & \textbf{0.2437} & 0.3318 & \textbf{0.1796} &0.2095 \\

\bottomrule
\end{tabular}
\label{tab:main}
\end{table*}


\subsection{Model Training}

Bayesian Personalized Ranking (BPR) loss is commonly uesd for primary recommending prediction follows:
\begin{equation}
    \label{eq:lossRec}
    \mathcal{L}_{rec}=\sum_{{(u,v^+,v^-)\in \Omega}}{-\log{\sigma(\hat{y}_{uv^+}-\hat{y}_{uv^-})}},
\end{equation}
where \(\Omega=\{(u,v^+,v^-)|{(u,v^+)}\in\Omega^{+},{(u,v^-)}\in\Omega^{-}\}\) represents the training set of triplet data, where \(\Omega^{+}\) denotes observed interactions and \(\Omega^{-}\) denotes the unobserved ones. \(\hat{y}\) indicates the users' preference score for items.


Totally for model training, \(\lambda_c\),\(\lambda_d\) and \(\lambda_r\) are hyperparameters that respectively control the strengths of contrastive learning, embedding dispersion and original prediction. In all, the performance of recommendation predictions are updated by optimizing this loss function:
\begin{equation}
    \label{eq:lossOverall}
    \mathcal{L}=\mathcal{L}_{rec}+\lambda_c(\mathcal{L}_{cl}^{(U)}+\mathcal{L}_{cl}^{(I)})\\
    +\lambda_d\mathcal{L}_{disp}+\lambda_r{||\Theta||}_F^2,
\end{equation}
where \({||\Theta||}_F^2\) denotes an L2 regularization term with a low weight \(\lambda_r\). 

\section{Experiments}

\label{sec:eva}

\subsection{Experimental Settings}

\paragraph{Evaluation Datasets}
For convincing results, we have conducted experiments using two widely recognized real-world datasets: Yelp\footnote{https://www.yelp.com/dataset} and Last.FM\footnote{https://www.last.fm/api}. The parameter details of these datasets are presented in Table \ref{tab:dataset}.

\textbf{Yelp}: A commonly utilized dataset encapsulates users' rating interaction collected on the Yelp platform, which allows users to share their check-ins about local venues.

\textbf{Last.FM}: A dataset collected from an online music radio platform, containing information such as tagging, social networking, and music preferences, etc.

\paragraph{Evaluation Metrics}
We employ two widely-used metrics to assess the prediction accuracy of all implemented methods: Recall@N and Normalized Discounted Cumulative Gain NDCG@N, which are computed by the all-ranking protocol \cite{LightGCN-He2020LightGCNSA}. Recall@N quantifies the correctness of identifying items within top-N list derived from ground truth, and NDCG@N gives a higher score to better ranking positions. 



\paragraph{Hyperparameter Settings}
For model inference, we optimize the learning process by employing the Adam optimizer and set the learning rate to \(1e^{-3}\), and the decay ratio to \(0.96\). Dimensionalities of bilateral embeddings are configured as 32. In our experiments, we adjust the quantity of hyperedges of models combined with hypergraph structure are set following the original paper. The regularization weights \(\lambda_c,\lambda_d\) and \(\lambda_r\) are taking values from the range \({ \{  1e^{-2}, 1e^{-1}, 1, 10, 100\}}\) for loss balance. The temperature parameters \(\tau_c\) and \(\tau_d\) are searched from the set \( \{ { 1e^{-2}, 1e^{-1}, 1, 10}\}\) to regulate the intensity of the gradients in our contrastive learning process. The number of convolutional layers that are set to three as best.

\subsection{Recommandation Performance}

We evaluate the effectiveness of BusGCL on a unified SSL recommendation framework called  SSLRec \cite{SSLRec-ren2023sslrec}, which achieves a more fair performance evaluation based on unified source data processing and sampling. The results are summarized in Table \ref{tab:overall}. Due to taking self-supervising training to fill the data gap in recommendation, some methods such as SGL \cite{SGL-Wu2020SelfsupervisedGL} and SimGCL \cite{SimGCL-Yu2021AreGA}, outperform earlier approaches like NGCF \cite{NGCF-10.1145/3331184.3331267} and LightGCN \cite{LightGCN-He2020LightGCNSA} with metric Recall@40 and NDCG@40 in Last.FM dataset especially. However, when comparing with all baselines, BusGCL not only achieves higher performance in two datasets but also demonstrates superiority over other existing methods in the Yelp dataset. Due to pairing subviews considerately, BusGCL can gather structural-similar user nodes properly and fit the relatively dispersed item nodes.  



\begin{table}[]
\centering
\caption{Ablation study on different BusGCL variants with/without dispersing loss. The metrics use Recall@20 and NDCG@20. ``Disper.'' is short for dispersing loss. }
\label{tab:ablation}
\begin{tabular}{lccccc}
\toprule
\multirow{2}{*}{Variants} & \multirow{2}{*}{Disper.} & \multicolumn{2}{c}{Yelp} & \multicolumn{2}{c}{Last.FM} \\
\cmidrule(r){3-4}
\cmidrule(r){5-6}
& & Recall  & NDCG  & Recall  & NDCG          \\
\midrule
BusGCL\(_{per}\)   & ×           & 0.0713 & 0.0355     & 0.2279 & 0.1711    \\
                   &\checkmark   & 0.0781 & 0.0388   & 0.2295 & 0.1718  \\
                   \midrule
BusGCL\(_{hyp}\)   & ×       & 0.0822 & 0.0415    & 0.2342 & 0.1742 \\
                   &\checkmark  & 0.0824 & 0.0417    & 0.2329 & 0.1728  \\
                   \midrule
BusGCL\(_{rev}\)   & ×      & 0.0817 & 0.0408     & 0.2236 & 0.1663   \\
                   &\checkmark  & 0.0827 & 0.0416 & 0.2234 & 0.1662   \\
                   \midrule
BusGCL             & ×   & 0.0824 & 0.0417   & 0.2436 & 0.1796\\
                   &\checkmark   & 0.0840 & 0.0424    &  0.2439 &  0.1797  \\
    
\bottomrule
\end{tabular}
\end{table}

\subsection{Ablation Experiment}
 To investigate the impact of different selections about subview and dispersing loss, we conduct the results of the BusGCL on two datasets for three subview selections with/without dispersing loss. The performance of each variant is shown in Table \ref{tab:ablation}. Variants named $X_{hyp} \/ X_{per}$ refers to that the two slices of contrastive views are both from hypergraph-/perturbing-GCN, and $X_{rev}$ means the selection of slices is reversed to BusGCL. Variants marked w/o disp means that the dispersing loss module is disabled. The Analysis is separate as follows:


\subsubsection{\textbf{Ablation of subview combination}} By comparing all kinds of combinations of contrastive views generated from hypergraph- and perturbing-GCN, BusGCL$_{hyp} $ and BusGCL$_{rev}$ gains close result to BusGCL, which increases accuracy by $ 0.5\%$ than BusGCL$_{per}$. And it is obvious that BusGCL has the best performance because of its tendency to seek aggregation on the user-side by hypergraph-GCN and retain differences on the item-side thought perturbing-GCN, which better reflects the real-world situation.

\subsubsection{\textbf{Ablation of dispersing loss}} Observing the impact of dispersing constrain on related subview combinations vertically, it is explicit that dispersing loss has a more significant improvement on recommendation performance where the combining view is generated by perturbing-GCN like BusGCL$_{per}$. This fact proves the effectiveness of dispersing in overcoming the over-smooth problem among nodes caused by the introduction of random noise.

\begin{figure}[t]
\centerline{\includegraphics[scale=0.28]{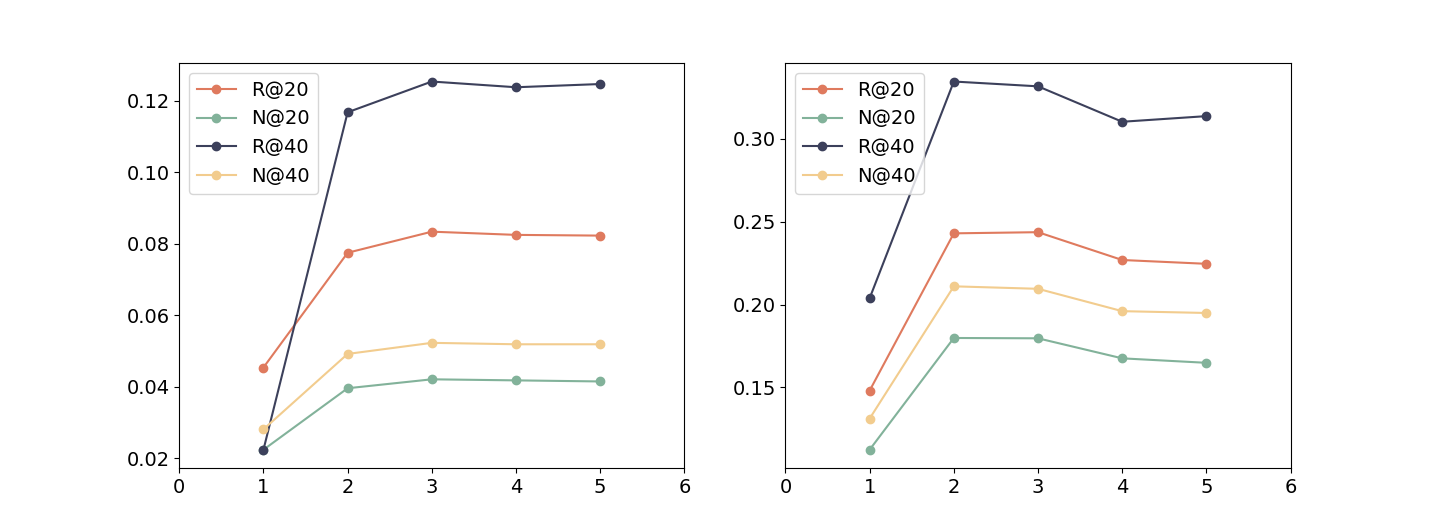}}
\caption{The effect of GCN layers, the data of the two images were measured on the Yelp and Last.FM datasets, respectively.}
\label{fig:layer}
\end{figure}

\subsection{Further Visualization and Analysis}


\subsubsection{ \textbf{The influence on hyperparameter of GCN layers} }To analyze the impact of varying the number of GCN layers, we initialize it with values in the set \(\{1, 2, 3, 4, 5\}\). The outcomes can be visualized in Figure \ref{fig:layer}. Upon reaching a total of \(3\) GCN layers, the model attains its most optimal performance on the Yelp dataset, and on Last.FM, models with $2$ or $3$ layers both have relatively good performance. As the number of layers continues to increase, the performance on both datasets decreases, the phenomenon we attribute to the prevalent issue of over-smoothing positively correlating with model depth.

\subsubsection{\textbf{The impact of hyperparameters of dispersing loss}}
\begin{figure}[t]
\centerline{\includegraphics[scale=0.27]{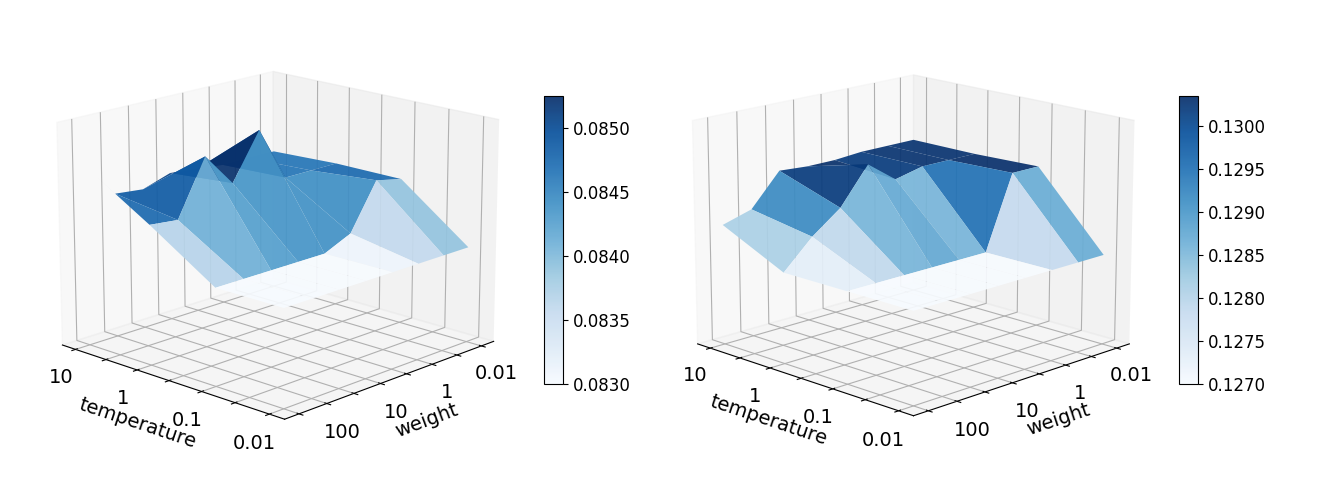}}
\caption{Influence of weight and temperature coefficient about dispersing loss on recommendation performance on Yelp dataset, that measurement by Recall@20 is presented in the left image and NDCG@20 on the right.}
\label{fig:surface}
\end{figure}

Figure \ref{fig:surface} shows the synergistic effect of temperature coefficient \(\tau_d\) and weight hyperparameter \(\lambda_d\) about dispersing loss. The strength of the temperature coefficient \(\tau_d\) of \(\mathcal{L}_{disp}\) affects the constraint effect on the relation between embeddings by controlling the smoothness of logit. It can be observed that the optimal performance is attained when \(\tau_d=1.0\). A lower \(\tau_d\) may lead to over-sensitivity of the disperse loss function, which causes the performance degradation. On the other hand, when the value of \(\lambda_d\) is set to 1, the model's recommendation performance reaches its peak.

\begin{figure*}[t]
\centerline{\includegraphics[scale=0.24]{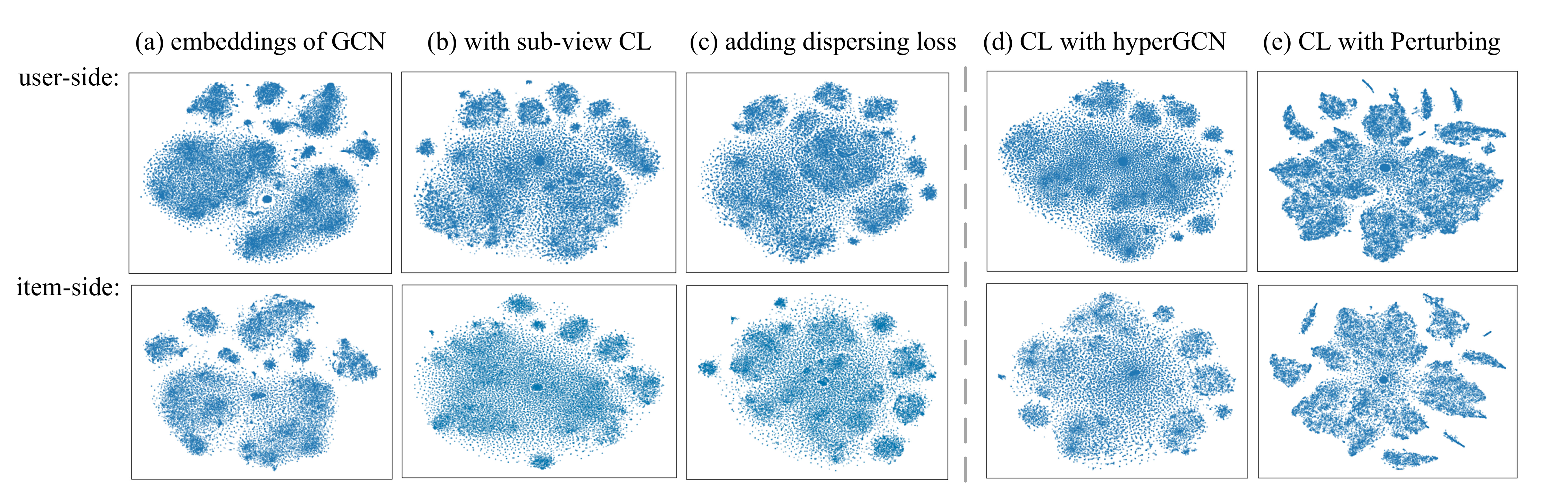}}
\caption{Visualization of distribution from different stages of BusGCL training by t-SNE on Yelp dataset. (a) shows embeddings after GCN process and readout directly. (b) is after bilateral subview CL. (c) adds dispersing loss. Additionally, (d) and (e) illustrate the situation when subviews on both sides are selected from hyperGCN/GCN with perturbing.}
\label{fig:five}
\end{figure*}

\subsubsection{ \textbf{Visualization results}}
We visualize the distribution of embeddings during the training process by t-SNE to seek depth analysis of how each module improves performance and their specific role in the embedded training process. As shown in Figure \ref{fig:five}, from graph (a) to (b), the blank areas between groups are diffused, which means CL makes the embedding distribution more uniform. This is because CL can find more implicit similarities between nodes, thereby reducing the gap between groups. Comparing (b) and (c), we can observe that on the relatively balanced distribution, several larger group structures have been formed, and separate the dense embeddings at the center to a certain extent. This means the dispersing loss helps to preserve differences between similar embeddings by generally pushing away all the other embeddings to more manifest subtle collaborative signals. 


By observing graph (d) and (e), it's not difficult to find that distributions of two-side embeddings produced by CL with symmetrical GCN structures are more similar in appearance, which makes it difficult to retain the inherent structural differences between the users and items. The final embedding visualization of our model is more suitable for data distribution in real world compared to embeddings of previous models.





\subsubsection{ \textbf{The impact of user-item model selection with different graph augmentations}}
To deeply analyze the rationality of using hyperGCN and GCN with perturbing as two graph augmentation methods in BusGCL, we compared multiple popular graph augmentations which are following \cite{SGL-Wu2020SelfsupervisedGL} on the subview-CL mechanism in BusGCL, such as edge dropout, node dropout and random walk. Based on the experimental results in Table \ref{tab:augs}, we can draw the following conclusion: Compared to other GCN models, the node drop-based model has fine performance. On the contrary, the effects of edge drop or random walk are not ideal. We believe that masking the adjacency matrix cannot guarantee the same distribution of the augmented graph and the original graph. Especially when calculating multi-hop relationships, this deviation will be amplified. The GCN combination of hypergraphs and perturbing we selected has the best experimental performance, which verifies their aforementioned adaptability on user- and item-side, respectively. Our model takes the bilateral inter-node characteristics into account when selecting graph augmentation methods, and makes efforts in preserves the differences between users and items.



\begin{table}[]
\centering
\caption{The impact of different GCN model selections with different graph augmentations on Yelp dataset. ``Hyper.'' means HyperGCN.}
\label{tab:augs}
\begin{tabular}{cccc}
\toprule
User model & Item model &  Recall@20  & NDCG@20        \\
\midrule
Hyper. & Perturb  & 0.0840 & 0.0424  \\
Hyper. & Node drop & 0.0819 & 0.0419\\
Hyper. & Edge drop & 0.0509 & 0.0266\\
Hyper. & Random walk &  0.0655 & 0.0336 \\
Node drop & Perturb & 0.0820& 0.0410\\
Edge drop & Perturb & 0.0429 & 0.0220\\
Random walk & Perturb & 0.0594 & 0.0302\\
    
\bottomrule
\end{tabular}
\end{table}

\subsubsection{ \textbf{The effect of dispersing loss compare to others}}
To further verify the suitability of dispersing loss in the BusGCL framework, we attempt to replace the dispersing loss functions with another Kullback-Leibler divergence based loss function that can constrain the embedding probability distribution and experiment with its effectiveness. 
The results with optimized loss weights link to Table \ref{tab:losses}. Specifically, we calculate the KL divergence of the bilateral embedding matrix for uniformly distributed matrices of the same shape as the loss value. 

Compared with the variant without additional losses, the addition of KL divergence and dispersing loss both results in an increase in performance, indicating the significance of constraining the embedding distribution, that making the embedding more evenly distributed in the vector space does indeed help to improve the ability of contrastive learning. Our dispersing loss outperforms KL-divergence, indicating the embedding constraints based on positive and negative sample metrics are more suitable in the BusGCL framework. Essentially, dispersing loss is a trick used in the model training process, and its involvement enhances the ability of learning implicit collaborative relationship between embeddings.



\begin{table}[]
\centering
\caption{The impact of alternative loss selections for mutual training in user-item reasoning.}
\begin{tabular}{ccccc}
\toprule
\multirow{2}{*}{Loss } & \multicolumn{2}{c}{Yelp} & \multicolumn{2}{c}{Last.FM} \\
\cmidrule(r){2-3}
\cmidrule(r){4-5}
&  Recall@20  & NDCG@20  & Recall@20  & NDCG@20          \\
\midrule

Dispersing loss  &  0.0840& 0.0424 & 0.2437 & 0.1796 \\
KL-divergence & 0.0830 & 0.0421 & 0.2425 & 0.1794 \\
No loss & 0.0824 & 0.0417 & 0.2402 & 0.1771\\

\hline
\end{tabular}
\label{tab:losses}
\end{table}


\section{Conclusion}
\label{sec:con}

In this paper, we improve the recommendation system by utilizing the bilateral unsymmetry of node density on the user- and item-side, and propose bilateral slicing contrastive learning which generates user and item subviews to reason better results. Then we propose a multi-struct graph framework BusGCL, which considers the characteristics of different GCNs and from which we select bilateral subviews to match the relation density difference between two sides. For training, a dispersing loss is designed to alleviate the over-smoothing issue deteriorated by GCN. Overall experiments and extensive studies validate the superiority of our proposed framework BusGCL towards classic and competitive baselines because of its adaptability to real-world data. In future work, we may explore adaptive methods for contrastive view generation to expand the universality of graph contrastive learning for more specific recommendation tasks.

\section*{Acknowledgment}
This work was supported in part by the National Natural Science Foundation of China under Grant 62372335.

\label{sec:ack}


\printbibliography
\end{document}